\documentclass[a4paper,fleqn,usenatbib]{mnras}

\usepackage{newtxtext,newtxmath}

\usepackage[T1]{fontenc}
\usepackage{ae,aecompl}

\usepackage{xcolor}
\usepackage{placeins}
\usepackage{tikz}

\usepackage{graphicx}	
\usepackage{amsmath}	
\usepackage{amssymb}	
\usepackage{array}

\title[Potential crust cooling of 1RXS J1804]{Potential cooling of an accretion-heated neutron star crust in the low-mass X-ray binary 1RXS J180408.9$-$342058}

\author[A.S. Parikh et al.]
{\parbox{\textwidth}{
A.S. Parikh$^{1}$\thanks{E-mail: a.s.parikh@uva.nl},
R. Wijnands$^{1}$,
N. Degenaar$^{1,2}$,
L.S. Ootes$^{1}$,
D. Page$^{3}$,
D. Altamirano$^{4}$, 
E.M. Cackett$^{5}$, 
A.T. Deller$^{6}$,
N. Gusinskaia$^{1}$,
J.W.T. Hessels$^{1,6}$,
J. Homan$^{7}$,
M. Linares,$^{8,9,10}$,
J.M. Miller$^{11}$
and J.C.A. Miller-Jones$^{12}$}
\vspace{0.4cm}\
\\
$^{1}$Anton Pannekoek Institute for Astronomy, University of Amsterdam, Postbus 94249, 1090 GE Amsterdam, The Netherlands\\
$^{2}$Institute of Astronomy, University of Cambridge, Madingley Road, Cambridge, CB3 OHA, UK\\
$^{3}$Instituto de Astronom\'{i}a, Universidad Nacional Aut\'{o}noma de M\'{e}xico, Mexico D.F. 04510, Mexico\\
$^{4}$Department of Physics and Astronomy, Southampton University, Southampton SO17 1BJ, UK\\
$^{5}$Department of Physics and Astronomy, Wayne State University, 666 W. Hancock St, Detroit, MI 48201, USA\\
$^{6}$ASTRON, the Netherlands Institute for Radio Astronomy, Postbus 2, 7990 AA, Dwingeloo, The Netherlands\\
$^{7}$Massachusetts Institute of Technology, Kavli Institute for Astrophysics and Space Research, Cambridge, MA 02139, USA\\
$^{8}$Instituto de Astrof\'{i}sica de Canarias, c/ V\'{i}a L\'{a}ctea s/n, E-38205 La Laguna, Tenerife, Spain\\
$^{9}$Universidad de La Laguna, Departamento de Astrof\'{i}sica, E-38206 La Laguna, Tenerife, Spain\\
$^{10}$Institutt for fysikk, NTNU, 7491 Trondheim, Norway\\
$^{11}$Department of Astronomy, University of Michigan, 500 Church Street, Ann Arbor, MI 48109, USA\\
$^{12}$International Centre for Radio Astronomy Research, Curtin University, GPO Box U1987, Perth, WA 6845, Australia
}

\date{Accepted XXX. Received YYY; in original form ZZZ}

\pubyear{2016}

\begin{document}
\label{firstpage}
\pagerange{\pageref{firstpage}--\pageref{lastpage}}
\maketitle

\begin{abstract}
We have monitored the transient neutron star low-mass X-ray binary 1RXS J180408.9$-$342058 in quiescence after its $\sim$4.5 month outburst in 2015. The source has been observed using {\it Swift} and {\it XMM-Newton}. Its X-ray spectra were dominated by a thermal component. The thermal evolution showed a gradual X-ray luminosity decay from $\sim$18$\times 10^{32}$ to $\sim$4$\times 10^{32}$ ($D$/5.8 kpc)$^{2}$ erg s$^{-1}$ between $\sim$8 to $\sim$379 days in quiescence and the inferred neutron star surface temperature (for an observer at infinity; using a neutron star atmosphere model) decreased from $\sim$100 to $\sim$71 eV. This can be interpreted as cooling of an accretion-heated neutron star crust. Modeling the observed temperature curve (using \texttt{NSC\textsc{ool}}) indicated that the source required $\sim$1.9 MeV per accreted nucleon of shallow heating in addition to the standard deep crustal heating to explain its thermal evolution. Alternatively, the decay could also be modelled without the presence of deep crustal heating, only having a shallow heat source (again $\sim$1.9 MeV per accreted nucleon was required). However, the {\it XMM-Newton} data statistically required an additional power-law component. This component contributed $\sim$30 per cent of the total unabsorbed flux in 0.5 -- 10 keV energy range. 
The physical origin of this component is unknown. One possibility is that it arises from low-level accretion. The presence of this component in the spectrum complicates our cooling crust interpretation because it might indicate that the smooth luminosity and temperature decay curves we observed may not be due to crust cooling but due to some other process.
\end{abstract}




\section{Introduction}
Transient low-mass X-ray binaries (LMXBs) harbouring neutron stars are excellent laboratories to probe the neutron star structure. The donors in LMXBs are typically $\lesssim 1\, M_{\odot}$ and mass transfer on to the primary is by Roche lobe overflow of the donor. Transient neutron star LMXBs experience periods of accretion outbursts ranging from weeks to years. These outbursts can reach X-ray luminosities of $L_X \sim$10$^{35 - 38}$ erg s$^{-1}$. 
Between the outbursts are periods of quiescence, which can last for a period of months to decades and the systems then typically have luminosities of $L_X \sim$10$^{30 - 34}$ erg s$^{-1}$. 

During outbursts the accreted material  compresses the neutron star crust and heats it up by electron capture, neutron emission, and density driven fusion reactions \citep[e.g.,][]{haensel1990non,haensel2008models,steiner2012deep} and the crust can be heated out of equilibrium with the core if enough heat is generated. In total about 1.5 -- 2 MeV per accreted nucleon is expected to be released deep in the crust (primarily at densities of 10$^{12 - 13}$ g cm$^{-3}$; \citeauthor{haensel2008models} \citeyear{haensel2008models}). When accretion ceases in quiescence the crust cools down mainly by conducting the heat to the core and the surface.  The thermal evolution of the neutron star crust is determined by the structure and composition of the crust, the strength and depth of the heating processes, and the properties of the accretion outburst \citep[e.g.,][]{shternin2007neutron,brown2009mapping,page2013forecasting}. The outermost layers cool fastest and hence as time progresses we probe the cooling of deeper and deeper layers.

So far eight transient neutron star LMXBs have been monitored in quiescence to study crust cooling \citep{wijnands2001chandra,wijnands2002xmm,wijnands2003chandra,wijnands2004monitoring,
cackett2006cooling,cackett2008cooling,cackett2010continued,cackett2013change,
degenaar2009chandra,degenaar2011evidence,degenaar2013continued,degenaar2014probing,
degenaar2015neutron,degenaar2011soft,degenaar2011accretion,
fridriksson2010rapid,fridriksson2011variable,trigo2011xmm,homan2014strongly,waterhouse2016constraining,merritt2016neutron}. Reconstructing the obtained results using various theoretical cooling models has provided us with new insights into the properties of neutron star crusts \citep{rutledge2002crustal,shternin2007neutron,brown2009mapping,page2013forecasting,medin2015time,horowitz2015disordered,turlione2015quiescent,deibel2015strong,ootes2016}. The rapid temperature decay that the cooling crust exhibits implies a very high thermal conductivity. The thermal conductivity is parametrized by the level of atomic impurities in the crust, $Q_\mathrm{imp}$ , where a lower level of impurities implies a higher thermal conductivity 

In addition, the high crust temperature in some systems can only be compatible with the theoretical models if an extra source of heat at relatively shallow depths is present which is not predicted by the standard deep crustal heating model described above. Other systems do not require this shallow heating to explain their thermal evolution \citep{page2013forecasting,degenaar2015neutron}. For those that need shallow heating the magnitude of the heating is not the same for all. Several sources studied need $\sim$1 -- 2 MeV per accreted nucleon \citep{brown2009mapping,degenaar2011evidence,degenaar2014probing,waterhouse2016constraining}, however, MAXI J0556$-$332 requires an exceptionally high magnitude of shallow heating of $\sim$6 -- 10 MeV per accreted nucleon \citep{deibel2015strong}. The cause of the shallow heating is unknown; we refer to the in-depth discussions in \citet{degenaar2013direct} and \citet{deibel2015strong} for more information about the potential shallow heating processes and the uncertainties therein.

\subsection{1RXS J180408.9$-$342058}
1RXS J180408.9$-$342058 (hereafter 1RXS J1804) was first detected by {\it ROSAT} in 1990 \citep{voges1999rosat}. It was detected again by {\it Integral} and {\it Swift} in 2012 April during a faint accretion outburst (the source was only detected at luminosities of $L_X \sim$10$^{34}$ erg s$^{-1}$; \citeauthor{chenevez2012integral} \citeyear{chenevez2012integral}). It was classified as a neutron star LMXB as it showed a thermonuclear type-I burst during the {\it Integral} observation, reported by \citet{chenevez2012integral}. They calculated an upper limit on its distance as determined from the brightness of the thermonuclear burst, of $D$ $\lesssim$ 5.8 kpc (assuming an Eddington luminosity limit for helium-rich material $L_\mathrm{Edd}$ = 3.8$\times$10$^{38}$ erg s$^{-1}$; \citeauthor{kuulkers2003photospheric} \citeyear{kuulkers2003photospheric}). It subsequently returned to quiescence as reported by \citet{kaur2012swift}. It went into outburst again on 2015 January 22 as determined using the {\it BAT} instrument on board {\it Swift} \citep{barthelmy2015swift,barthelmy2015trigger,krimm2015swift}. The new outburst was also confirmed using {\it MAXI}/GSC \citep{negoro2015maxi}. The source showed a peak luminosity of $\sim$0.12 $L_\mathrm{Edd}$. A state transition from hard to soft was observed $\sim$70 days into the outburst \citep{degenaar2015neutron,degenaar2016disk}. Based on spectral fits that include relativistic disk reflection models, high-resolution X-ray studies of the outburst properties indicated that the accretion disk likely extended close to the neutron star surface both during the hard and soft spectral states \citep{ludlam2016nustar,degenaar2016disk}. 1RXS J1804 was also observed in the radio \citep{deller2015radio,gusinskaia2016} suggesting the presence of a jet. \citet{baglio20161rxs} reported on the NIR/optical/UV observations of the source during outburst which also suggests the presence of an jet. The source transitioned back to quiescence in 2015 June after a $\sim$4.5 month outburst.

In this work we present the quiescent X-ray observations from 1RXS J1804 after the end of its 2015 outburst. We report on possible crust heating and cooling of the neutron star in this source and model the cooling curve to obtain insight into the neutron star crust.

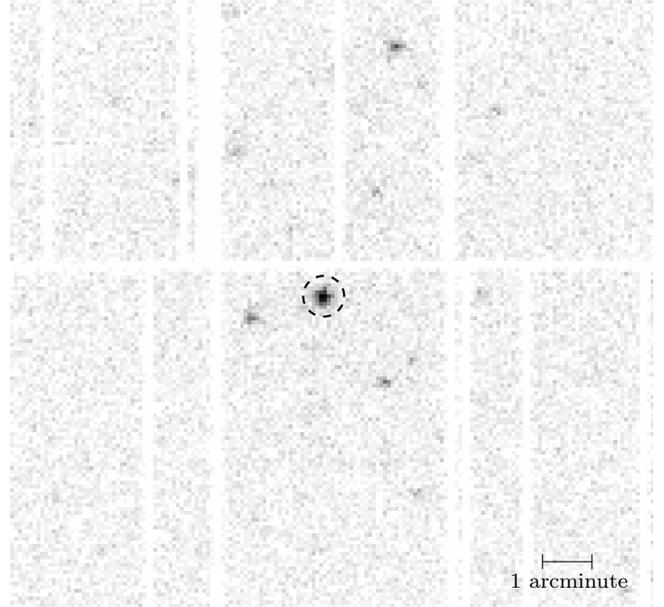
\begin{figure}
\centering
\begin{tikzpicture}
	\node[anchor=south west,inner sep=0] at (0,0) {\includegraphics[scale=0.24]{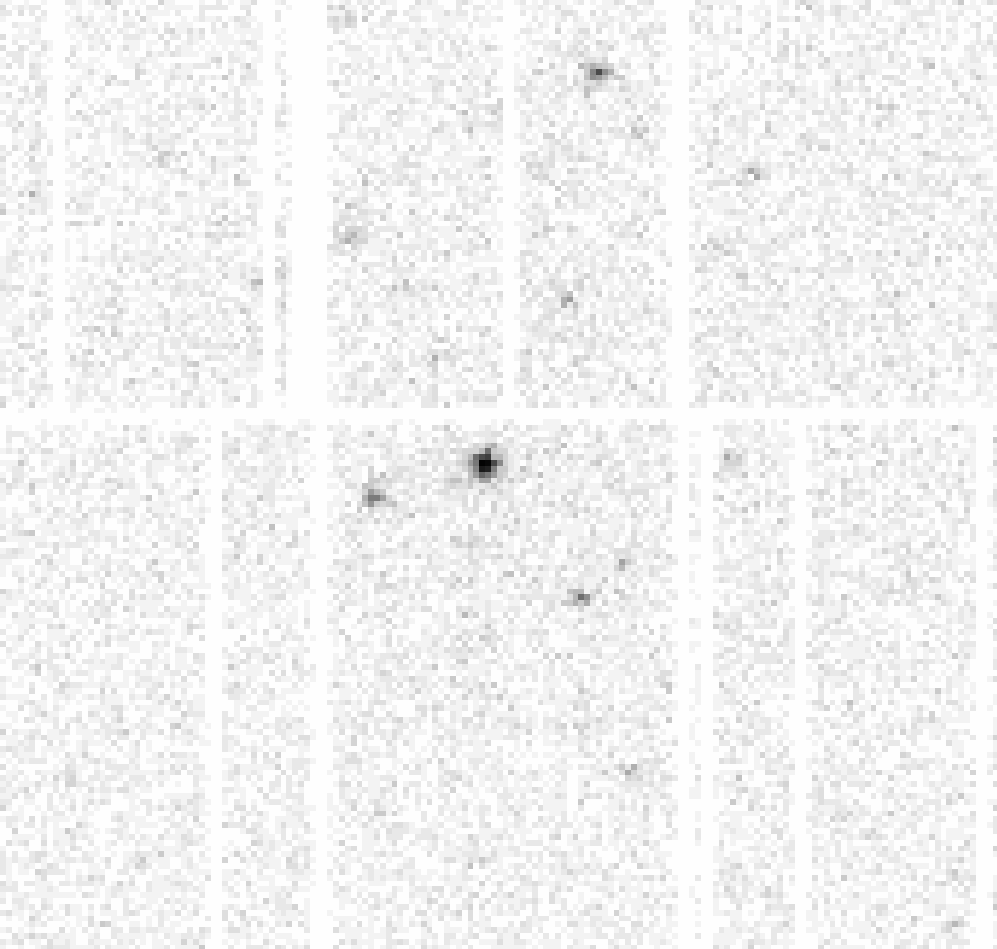}};
    \draw[black,thick,dashed](4.12,4.12) circle (0.27cm);
    \draw (7.65,0.6) -- (7,0.6) ;
     \draw (7.65,0.7) -- (7.65,0.5);
         \draw (7,0.7) -- (7,0.5);
 \node at (7.35,0.35) {1 arcminute};
\end{tikzpicture}
\caption{The image of the field near 1RXS J180408.9$-$342058 (indicated by the dashed circle of radius 25$"$) as obtained using the EPIC-PN camera on board {\it XMM-Newton}.}
\label{imag_src_pn}
\end{figure}

\begin{figure}
\centering
\includegraphics[scale=0.435]{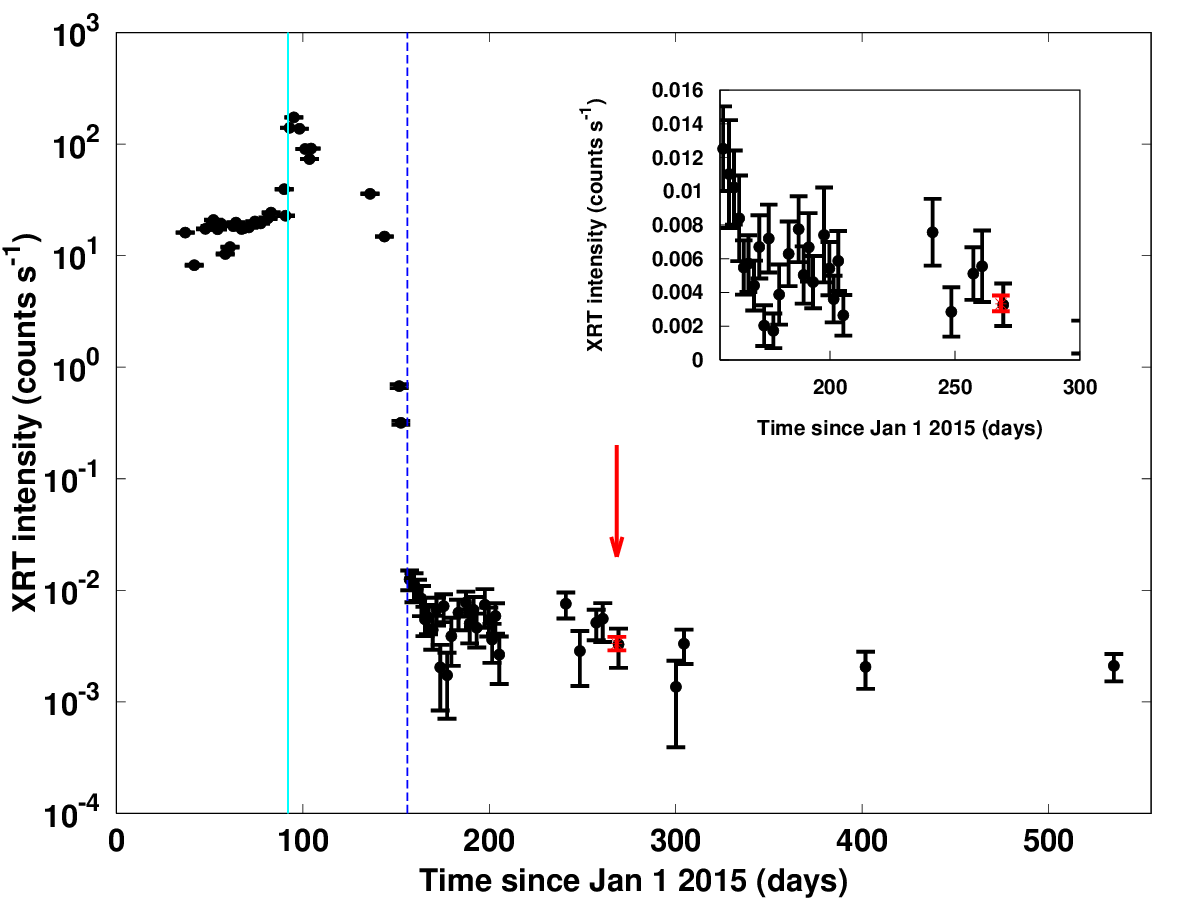}
\caption{The {\it Swift}/XRT light curve of 1RXS J1804 during its 2015 outburst and subsequent quiescence. The count rates are for the 0.5 -- 10 keV energy range and are averaged over each observation ID. The red arrow indicates the time of the {\it XMM-Newton} observation and the red $\ast$ indicates the {\it XMM-Newton} count rate converted to the {\it Swift}/XRT count rate using \texttt{WebPIMMS} (https://heasarc.gsfc.nasa.gov/cgi-bin/Tools/w3pimms/w3pimms.pl). The source transitioned from the hard state to the soft state around 2015 April 3, as shown by the solid cyan line. The transition into quiescence was estimated to be 2015 June 6, indicated by the dashed blue line. The inset shows a zoom of the count rate variation soon after the transition to quiescence.}
\label{img_1rxs_lightcurve}
\end{figure}

\section{Observations and Data Analysis}
We monitored the outburst, subsequent decay, and quiescence of 1RXS J1804 with observations performed using the X-ray Telescope (XRT) on board {\it Swift} \citep{burrows2005swift}. We also observed the source using {\it XMM-Newton} \citep{struder2001european,turner2001european} $\sim$112 days into quiescence. In Table \ref{tab_log_obs_1rxs} we show the log of the observations that were performed when the source was in quiescence. We also analyse an observation taken using {\it Chandra} in 2011 to check if the data could be used to estimate the base-level of the source in quiescence.

\begin{table}
\centering
\caption{Log of the observations used for 1RXS J1804 when it was in quiescence. The errors are calculated for the 1$\sigma$ confidence range.\textsuperscript{*}}
\label{tab_log_obs_1rxs}
\begin{tabular}{rccc}
\hline
Date & Observation & Exposure & Count Rate \tabularnewline
& ID & Time & (0.5 -- 10 keV) \tabularnewline
& & (ks)& (10$^{-3}$ counts s$^{-1}$) \tabularnewline
 \hline
 \multicolumn{4}{c}{  {\it Swift}/XRT} \tabularnewline
  \hline
  2015/06/07 &00033806002 & 2.0 & 12.6  $\pm$      2.5          \tabularnewline
 		09 &00032436037 & 1.1 & 11.1  $\pm$      3.2       	 \tabularnewline
 		11 &00033806003 & 2.1 & 10.8 $\pm$       2.2       	\tabularnewline
 		13 &00033806004 & 1.3 &  8.4 $\pm$       2.5       	\tabularnewline
 		15 &00033806005 & 2.2 &  5.6 $\pm$       1.6       	\tabularnewline
 		17 &00033806006 & 2.1 &  5.8 $\pm$      1.7       	\tabularnewline
 		19 &00033806007 & 2.0 &  5.4 $\pm$       1.6       	\tabularnewline
 		21 &00033806008 & 1.9 &  6.8 $\pm$      1.9       	\tabularnewline
 		23 &00033806009 & 1.5 &  2.1 $\pm$      1.2       	 \tabularnewline
 		25 &00033806010 & 1.8 &  7.8 $\pm$      2.1       	 \tabularnewline
 		27 &00033806011 & 1.7 &  2.4 $\pm$      1.2       	 \tabularnewline
 		29 &00033806012 & 1.3 &  4.0 $\pm$      1.8       	 \tabularnewline
 	  07/03 &00033806013 & 1.7 &  6.3 $\pm$     1.9       	 \tabularnewline
 		07 &00033806015 & 2.0 &  7.8 $\pm$     2.0       	 \tabularnewline
 		09 &00033806016 & 1.8 &  5.7 $\pm$      1.8       	 \tabularnewline
 		11 &00033806017 & 1.7 &  6.7 $\pm$      2.0       	 \tabularnewline
 		13 &00033806018 & 2.0 &  4.6 $\pm$      1.5       	 \tabularnewline
 		17 &00033806020 & 1.0 & 7.4 $\pm$      2.8       	 \tabularnewline
 		19 &00033806021 & 2.2 &  5.9 $\pm$     1.6       	 \tabularnewline
 		21 &00033806022 & 1.9 &  3.7 $\pm$      1.4       	 \tabularnewline
 		23 &00033806023 & 1.9 &  5.9 $\pm$     1.8       	 \tabularnewline
 		25 &00033806024 & 1.9 &  4.8 $\pm$      1.6       	 \tabularnewline
 		30 &00032436038 & 2.0 &  7.7 $\pm$      2.0       	 \tabularnewline
 	  09/06 &00032436039 & 1.4 &  2.9 $\pm$      1.5       	 	\tabularnewline
 		15 &00032436040 & 2.1 &  5.6 $\pm$      1.6       	 	\tabularnewline
 		18 &00032436041 & 1.3 &  5.6 $\pm$     2.1      	 	\tabularnewline
 		27 &00032436042 & 2.1 &  3.8 $\pm$      1.3       	 	\tabularnewline
 	  10/28 &00032436043 & 1.5 &  2.8 $\pm$      1.4       		\tabularnewline
 	  11/01 &00032436044 & 2.7 &  3.4 $\pm$ 	 1.1		\tabularnewline
  2016/02/06 &00032436045 & 3.8 &  2.4 $\pm$      0.8       		\tabularnewline
  2016/06/19 &00032436046 & 6.4 &  2.3 $\pm$      0.9       		\tabularnewline
 \hline
\multicolumn{4}{c}{  {\it XMM-Newton}} \tabularnewline
  \hline
  2015/09/26 & 0770380301& 28.4 (MOS1)	&12.2 $\pm$ 0.7\tabularnewline
  				  & 					   & 29.3 (MOS2)	& 16.1 $\pm$ 0.9\tabularnewline
				  & 					   & 13.2 (PN)	& 46.2 $\pm$ 2.0\tabularnewline
\hline
\multicolumn{4}{p{8.4cm}}{\textsuperscript{*}\scriptsize{The source position for {\it Swift} Observation ID 00033806014 (2015 July 5) was located on a bad column and the source was not detected. Therefore, this observation has not been included in the analysis. {\it Swift} Observation ID 00033806019 (performed on 2015 July 15) and  00032436047 (performed on 2016 June 21) had short exposure times (0.35 ks and 0.5 ks respectively). Since the count rate upper limit is rather unconstraining we do not include these observation in our analysis.}} \tabularnewline
\end{tabular}
\end{table}

\subsection{\textit{\textbf{Swift}}/XRT}
1RXS J1804 has been frequently monitored using {\it Swift} except during the Solar constraint window between November 2015 and February 2016. The most recent observation of the source (00032436047) was taken on 2016 June 21. 

{\it Swift} was used to observe 1RXS J1804 for a total of 34 times in quiescence (see Table \ref{tab_log_obs_1rxs}). All the quiescent data were collected with the XRT operating in Photon Counting mode. We obtained the data from the \texttt{HEASARC} archive\footnote{http://swift.gsfc.nasa.gov/archive/} and analysed it using \texttt{HEASOFT} (version 6.17). The raw data were processed using \texttt{xrtpipeline} and \texttt{XS\textsc{elect}} (version 2.4c) was used for light curve and spectra extraction. A circular source extraction region of radius 25$"$ was used. 
 The {\it XMM-Newton} observation (see Figure \ref{imag_src_pn} and Section \ref{sect_xmm_newton}) shows several other X-ray sources in the 1RXS J1804 field of view so an annular extraction region could not be used for the background correction. Therefore, for the background extraction region we used a circular region of radius 75$"$ placed on a source free part of the CCD (as determined from the {\it XMM-Newton} image shown in Figure \ref{imag_src_pn}). 
The tool \texttt{xrtexpomap} was used to correct for faulty pixels/columns. The ancillary response files were generated using \texttt{xrtmkarf} and the appropriate response matrix files, as indicated by \texttt{xrtmkarf}, were used.

The light curve for the 2015 outburst and subsequent quiescence of 1RXS J1804 is shown in Figure \ref{img_1rxs_lightcurve}. The count rates were averaged over each observation ID and 6 thermonuclear type-I bursts 
were removed for the count rate calculation. The count rates were corrected for the background (but not corrected for faulty pixels/columns) and were determined for the 0.5 -- 10 keV energy range. Figure \ref{img_1rxs_lightcurve} also shows the subsequent transition of 1RXS J1804 into quiescence around 2015  June 6. This transition date was determined by the intercept of two lines -- an exponential was fit to the steep decay trend immediately before the end of the outburst and a straight line was fit to the shallower decay trend after transition into quiescence (the same method was also used by \citeauthor{fridriksson2010rapid} \citeyear{fridriksson2010rapid}). In this paper we are interested in following the spectral evolution of the neutron star in quiescence so from now on we will focus on quiescent data (the data to the right of the dashed blue line shown in Figure \ref{img_1rxs_lightcurve}), although the outburst data will be used for modeling the crust cooling curve of the source (see Section \ref{sect_crust_cooling_model}).

\subsection{\textit{\textbf{XMM-Newton}}}
\label{sect_xmm_newton}
1RXS J1804 was observed using {\it XMM-Newton} on 2015 September 26 at 9:21 UT (see Table \ref{tab_log_obs_1rxs} and Figure \ref{img_1rxs_lightcurve}). All three EPIC detectors -- MOS1, MOS2, and PN -- were operating in full window mode. The source was too faint to be detected using the RGS and therefore we do not discuss those data further. The data were downloaded from the {\it XMM-Newton} Science Archive\footnote{http://www.cosmos.esa.int/web/xmm-newton/xsa} (\texttt{XSA}) and were reduced using the Science Analysis System (\texttt{SAS}; version 14.0) and the raw data were processed using \texttt{emproc} and \texttt{epproc} for the MOS and PN detectors respectively. We searched for background flaring by examining the light curves for the energy range >10 keV for the 2 MOS detectors, and between 10 and 12 keV for the PN detector. We excluded those intervals from our data in which the count rate exceeded 0.2 counts s$^{-1}$ for the MOS1, 0.25 counts s$^{-1}$ for the MOS2, and 0.4 counts s$^{-1}$ for the PN. This resulted in an effective exposure time of 28, 29, and 13 ks for the MOS1, MOS2 and PN detectors, respectively. Spectral extraction was carried out using \texttt{xmmselect} with a circular source extraction region having a radius of 25$"$ (see Figure \ref{imag_src_pn}). A background extraction region of radius 75$"$, placed over a source free part of the same CCD, was used. These correspond to the same source and extraction regions used during the {\it Swift} analysis. We used \texttt{rmfgen} and \texttt{arfgen} to generate the redistribution matrix file and ancillary response function.

\subsection{\textit{\textbf{Chandra}}}
1RXS J1804 was observed using the {\it Chandra} X-ray observatory on 2011 July 24 for $\sim$4 ks (obs ID 12947) using the S3 chip of the Advanced CCD Imaging Spectrometer (ACIS; \citeauthor{garmire2003proc} \citeyear{garmire2003proc}) in the very faint mode, using the 1/8 sub-array to prevent possible pile-up. The data were downloaded from the {\it Chandra} archive\footnote{http://cda.harvard.edu/chaser/} and analysed using \texttt{CIAO} (version 4.8). The light curve extracted from the background region was inspected for possible flares. No background flares were detected so the all data were used for spectral analysis. Circular extraction regions of radius 2$"$ and 20$"$ were used for the source and background respectively. Both the extraction regions were placed on the same CCD.

\section{Results}

\begin{table*}
\centering
\caption{The intervals chosen to study the spectral evolution of 1RXS J1804.\textsuperscript{$\dagger$}}
\label{obsID_grouping}
\begin{tabular}{llllllll}
\hline
\multicolumn{8}{c}{  {\it Swift}/XRT} \tabularnewline
\hline
Interval& Time since&& $kT^{\infty}_{\mathrm{eff}}$ & && Flux & Luminosity\tabularnewline \vspace{0.8mm}  
&end of outburst&& (eV)& && (10$^{-13}$  &(10$^{32}$ erg s$^{-1}$)\tabularnewline
&(days)&&&& &erg cm$^{-2}$ s$^{-1}$) &\tabularnewline
\hline   \vspace{0.8mm} 
00033806002 -- 00033806008 \textsuperscript{$\star$}&8.3$_{-5.8}^{+4.9}$& & 100.0$^{+2.8}_{-3.2}$& &&4.5$_{-0.7}^{+0.8}$ & 18.0$_{-2.7}^{+3.0}$  \tabularnewline \vspace{0.8mm} 
00033806009 -- 00033806018&27.7$_{-8.0}^{+6.7}$ && 89.1$^{+3.3}_{-3.5}$& &&2.6$^{+0.6}_{-0.5}$&10.6$^{+2.3}_{-2.0}$\tabularnewline \vspace{0.8mm} 
00033806020 -- 00032436042&68.9$_{-24.6}^{+35.4}$ && 87.5$^{+3.1}_{-3.7}$&&&2.4$\pm$0.5&9.5$^{+2.1}_{-1.9}$\tabularnewline \vspace{0.8mm} 
00032436043 -- 00032436045&193.5$_{-35.6}^{+45.5}$ && 75.4$^{+6.3}_{-7.0}$&&&1.3$^{+0.7}_{-0.5}$&5.1$^{+2.7}_{-2.1}$\tabularnewline \vspace{0.8mm} 
00032436046 &379.0$\pm$0.7 && 71.3$^{+7.0}_{-7.9}$&&&1.0$^{+0.6}_{-0.5}$&3.7$^{+2.5}_{-1.8}$\tabularnewline
\hline
\multicolumn{8}{c}{  {\it XMM-Newton}} \tabularnewline

\hline
Observation ID&Time since&$N_\mathrm{H}$&  $kT^{\infty}_{\mathrm{eff}}$ &$\chi^2_{\upsilon}$/d.o.f. &$\Gamma$& Flux & Luminosity\tabularnewline
&end of outburst&(10$^{22}$ cm$^{-2}$)& (eV)&& &(10$^{-13}$& (10$^{32}$ erg s$^{-1}$) \tabularnewline 
&(days)&&&&&erg cm$^{-2}$ s$^{-1}$) &\tabularnewline 
\hline \vspace{0.8mm} 
0770380301&112.4$\pm$0.2&0.40$_{-0.04}^{+0.05}$ & 78.8$_{-6.7}^{+3.3}$& 1.14/59&1.9$_{-0.9}^{+1.0}$&2.3$\pm$0.1&9.0$\pm$0.5\tabularnewline \vspace{0.8mm} 
 && 0.4 (fixed) & 78.9$^{+2.3}_{-6.6}$& 1.12/60&1.9$^{+0.9}_{-0.8}$&2.3$\pm$0.1&9.1$\pm$0.5\tabularnewline \vspace{0.8mm} 
&  & 0.4 (fixed) & 78.7$^{+1.3}_{-1.4}$& 1.10/61&1.9 (fixed)&2.3$\pm$0.1&9.1$\pm$0.5\tabularnewline
\hline
\multicolumn{8}{p{18cm}}{\textsuperscript{$\dagger$}\scriptsize{For the {\it Swift} observations the $N_{\mathrm{H}}$ was fixed to 0.4$\times$10$^{22}$ cm$^{-2}$. $kT^{\infty}_{\mathrm{eff}}$ refers to the effective temperature for an observer at infinity (see footnote \ref{footnote_kT_inf}). Both the fluxes and the luminosities are for the 0.5 -- 10 keV energy range. The fluxes are the unabsorbed fluxes from all the contributing components. The luminosities are calculated assuming a distance of 5.8 kpc. The errors on the spectral parameters and the fluxes are calculated using the 90$\%$ confidence range using \texttt{XS\textsc{pec}}. For the {\it Swift} intervals the errors on time are determined using the weighted average of the exposure time of each observation present in a given interval. The time error bars for the {\it XMM-Newton} observation are determined from the exposure time of the observation. The data from the three {\it XMM-Newton} EPIC cameras have been fit simultaneously for each of the three different fits shown in the table.}}\tabularnewline
\multicolumn{8}{p{18cm}}{\textsuperscript{$\star$}\scriptsize{This interval includes the observation ID 00032436037.}}\tabularnewline

\end{tabular}
\end{table*}

\subsection{\textit{\textbf{Swift}}/XRT}
\label{sect_res_swift}
We combined several adjacent observations into `intervals' to obtain higher quality spectra. The intervals were chosen in such a way to reach a compromise between increasing the number of counts in the spectra and not averaging over too much of the spectral evolution that might occur between the different observations used in the intervals. The chosen intervals are shown in Table \ref{obsID_grouping}. The spectra were fit in \texttt{XS\textsc{pec}} (version 12.9; \citeauthor{arnaud1996xspec} \citeyear{arnaud1996xspec}). The obtained spectra still had only a small number of photons (a few tens to a maximum of around a hundred). The spectra were grouped using the \texttt{FTOOL} \texttt{grppha} to have at least 5 photons per bin. The spectra from the last two intervals had very few photons and were binned to contain a minimum of 1 photon per bin. Due to the low number of photons per bin we used W-statistics (background subtracted Cash statistics; \citeauthor{wachter1979parameter} \citeyear{wachter1979parameter}). The spectra could be fit with a variety of single component models. When using a power-law model the photon indices were very soft ($\Gamma$ = 2.4 -- 3.8) indicating a thermal-like spectrum. Therefore, we used a neutron star atmosphere model (i.e., \texttt{nsatmos}, \citeauthor{heinke2006hydrogen} \citeyear{heinke2006hydrogen}; relevant for neutron stars with a negligible magnetic field, $<10^9$ G) to fit our spectra. All parameters in the spectral model \texttt{nsatmos} were fixed, except the effective temperature. The equivalent hydrogen column density $N_\mathrm{H}$ was implemented using \texttt{tbabs}  with \texttt{WILM} abundances \citep{wilms2000absorption} and \texttt{VERN} cross-sections  \citep{verner1996atomic} throughout. Using different abundances may change the absolute values marginally \citep[see also][for a brief study of this effect; their Appendix A]{plotkin20162015} but the trend, and therefore the physical interpretation, would remain the same.

The equivalent hydrogen column density was set to $N_\mathrm{H}$ = 0.4$\times 10^{22}$ cm$^{-2}$ for all observations as this was the value obtained for the best-fit to the {\it XMM-Newton} data (see Section \ref{sect_analysis_powerlaw_res}). It appears that the source is viewed at a relatively low inclination of $\sim$20 -- 30 degrees (as has been determined using the reflection model fits; \citeauthor{ludlam2016nustar} \citeyear{ludlam2016nustar}, \citeauthor{degenaar2016disk} \citeyear{degenaar2016disk}) so assuming that the $N_\mathrm{H}$ remains constant appears reasonable. The distance was fixed at 5.8 kpc \citep{chenevez2012integral}. The mass and the radius of the neutron star was assumed to be the canonical $M_{\mathrm{NS}} = 1.6\,M_{\odot}$ and $R_{\mathrm{NS}} = 11$ km.\footnote{Tests using different $M_{\mathrm{NS}}$ and $R_{\mathrm{NS}}$ shows that although the absolute value of the inferred neutron star surface temperatures change slightly ($\sim$3 eV over the parameter space $M_{\mathrm{NS}} = 1.4 - 1.8\, M_{\odot}$ and $R_{\mathrm{NS}} = 10 - 12$ km) the physical interpretation remains the same.}  
The entire surface of the neutron star was assumed to be emitting (the normalization was set to 1). All the fluxes in this work were determined using the convolution model \texttt{cflux} in \texttt{XS\textsc{pec}}. The resulting spectral fits are shown in Table \ref{obsID_grouping} which indicates a steady decay in thermal flux and inferred neutron star surface temperature $kT^{\infty}_\mathrm{eff}$.\footnote{\label{footnote_kT_inf}The effective temperature as seen by an observer at infinity. $kT^{\infty}_\mathrm{eff}$ = $kT_\mathrm{eff}$/(1 + $z$), where 1 + $z$ is the gravitational redshift factor. For our assumed neutron star mass and radius of $M_{\mathrm{NS}} = 1.6 \,M_{\odot}$ and $R_{\mathrm{NS}} = 11$ km the gravitational redshift factor is 1 + $z$ = 1.32.} 

Since the {\it XMM-Newton} data required an additional power-law component (see Section \ref{sect_analysis_powerlaw_res}) the {\it Swift} data were re-analysed with a combined \texttt{nsatmos} and \texttt{powerlaw} model. To improve the constraints on model parameters the spectra from the five {\it Swift} intervals were fit simultaneously such that the value of the photon index was tied between the various intervals but the normalization was left free. Due to the low statistics of the data, the upper limits on any contribution by this added power-law component were very high and therefore unconstraining. We also tried to fix the photon index to $\Gamma$ = 1.9, the same value as the best-fit one found in the {\it XMM-Newton} data. Once again the upper limit on the power-law contribution was unconstraining. Thus, we could not exclude the presence of a power-law component similar to the one present in the {\it XMM-Newton} data.



\subsection{\textit{\textbf{XMM-Newton}}}
\label{sect_analysis_powerlaw_res}

\begin{figure}
\centering
\includegraphics[scale=0.34]{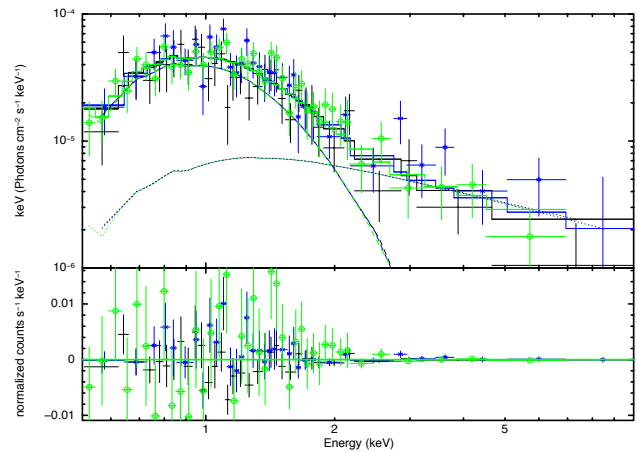}
\caption{The best-fit model for the {\it XMM-Newton} (MOS1, black +; MOS2, blue $\ast$; and PN, green $\circ$) data fit simultaneously is shown by the solid lines. The individual model components \texttt{nsatmos} and \texttt{powerlaw} are shown by the dashed and dotted lines respectively. For plotting purposes, a different rebinning has been used for the spectra than what was actually used in the spectral fits. The fit corresponds to the parameters shown in  Table \ref{tab_fit_NSCool}.}
\label{fig_xmm_model_fit}
\end{figure}

The spectra from the three {\it XMM-Newton} detectors (MOS1, MOS2, and PN) were simultaneously fit in \texttt{XS\textsc{pec}}. Using \texttt{specgroup} the spectra were grouped to contain at least 25 photons per bin to allow the use of the $\chi^2$ minimisation method during our spectral fits. At first, the equivalent hydrogen column density $N_\mathrm{H}$ was left free. Fitting the data with a power-law model revealed a very soft spectrum (photon index $\Gamma$ = 3.3), very similar to our {\it Swift} results. Therefore, we used an absorbed \texttt{nsatmos} model to fit the {\it XMM-Newton} spectrum. But this did not result in a satisfactory fit ($\chi^2_{\upsilon}$ = 1.90 for 61 d.o.f.). When including an additional power-law component a significantly better fit was obtained ($\chi^2_{\upsilon}$ = 1.14 for 59 d.o.f.). An $F$--test was used to calculate if the probability that this improvement of the fit, by adding the power-law component, was by chance. It gave a probability = 8.3$\times 10^{-8}$, showing that the extra power-law component was statistically required for an improved fit. The equivalent hydrogen column density, for the best-fit, was $N_\mathrm{H}$ = (0.40 $\pm$ $  $0.05)$\times 10^{22}$ cm$^{-2}$. This $N_\mathrm{H}$ is consistent with the values determined by \citet{krimm2015swift2} and \citet{degenaar2016disk}.

To produce results directly comparable to those from the {\it Swift} observation we fixed the equivalent hydrogen column density to $N_\mathrm{H}$ = 0.4$\times 10^{22}$ cm$^{-2}$. The best-fit (shown in Figure \ref{fig_xmm_model_fit}) revealed a photon index of $\Gamma$ = 1.9$^{+0.9}_{-0.8}$. When leaving the photon index free the resulting errors on the inferred neutron star surface temperature $kT^{\infty}_\mathrm{eff}$ were very large, as large or even larger than the errors obtained for the $kT_\mathrm{eff}^{\infty}$ in the {\it Swift} data despite the much higher quality of the {\it XMM-Newton} data. Therefore, to obtain more reasonable constraints on $kT^{\infty}_\mathrm{eff}$ we fixed the photon index to 1.9. The results of the various fits are tabulated in Table \ref{obsID_grouping}. The unabsorbed 0.5 -- 10 keV flux of the thermal component was $F$ = (1.5 $\pm$ 0.1)$\times10^{-13}$ erg cm$^{-2}$. The power-law component contributed $\sim$30 per cent to the total  unabsorbed 0.5 -- 10 keV flux. The power-law flux was $F$ = (7.0 $\pm$ 1.2)$\times10^{-14}$ erg cm$^{-2}$ s$^{-1}$  (0.5 -- 10 keV) and its luminosity was $L_X$ = (2.8 $\pm$ 0.5)$\times10^{32}$ ($D$/5.8 kpc)$^{2}$ erg s$^{-1}$.

We ran several tests to investigate if the presence of the additional power-law component was independent of our analysis technique. We processed the data by (1) using only `singles' events, (2) rejecting background flares more strongly, (3) using a different background region, and (4) only using 1/4 of the exposure time. Test (1), (2), and (3) still showed the presence of a statistically significant power-law component. This confirms that the presence of the additional power-law component in the {\it XMM-Newton} spectra is real and not an artefact of our data processing choices. In test (4) we manipulated the data to be of a lower-quality by only using 1/4 of the exposure time. For this test the power-law was not a statistically required additional component in the X-ray spectrum besides the \texttt{nsatmos} component. A single \texttt{nsatmos} fit to the X-ray spectrum ($N_\mathrm{H}$ fixed to 0.4$\times 10^{22}$ cm$^{-2}$) resulted in $kT^{\infty}_{\mathrm{eff}}$ = 78.9$^{+1.4}_{-1.7}$ ($\chi^2_{\upsilon}$ = 1.20 for 56 d.o.f.), which is very similar to the $kT^{\infty}_{\mathrm{eff}}$ determined from the original high quality data while an additional power-law component was used alongside \texttt{nsatmos}. This manipulated poor quality {\it XMM-Newton} data is roughly comparable to the quality of the {\it Swift} data. And we have found that fitting a single component \texttt{nsatmos} model to poor quality gives similar $kT^{\infty}_{\mathrm{eff}}$ to that from high quality data that requires an additional power-law component. This suggests that we can compare our $kT^{\infty}_{\mathrm{eff}}$ results from the {\it Swift} and {\it XMM-Newton} data directly in spite of the fact that we used two different types of models to fit the {\it Swift} and the {\it XMM-Newton} spectra.

\begin{figure}
\centering
\begin{tikzpicture}
\node[inner sep=0pt] (russell) at (0,0)
    {\includegraphics[scale=0.42]{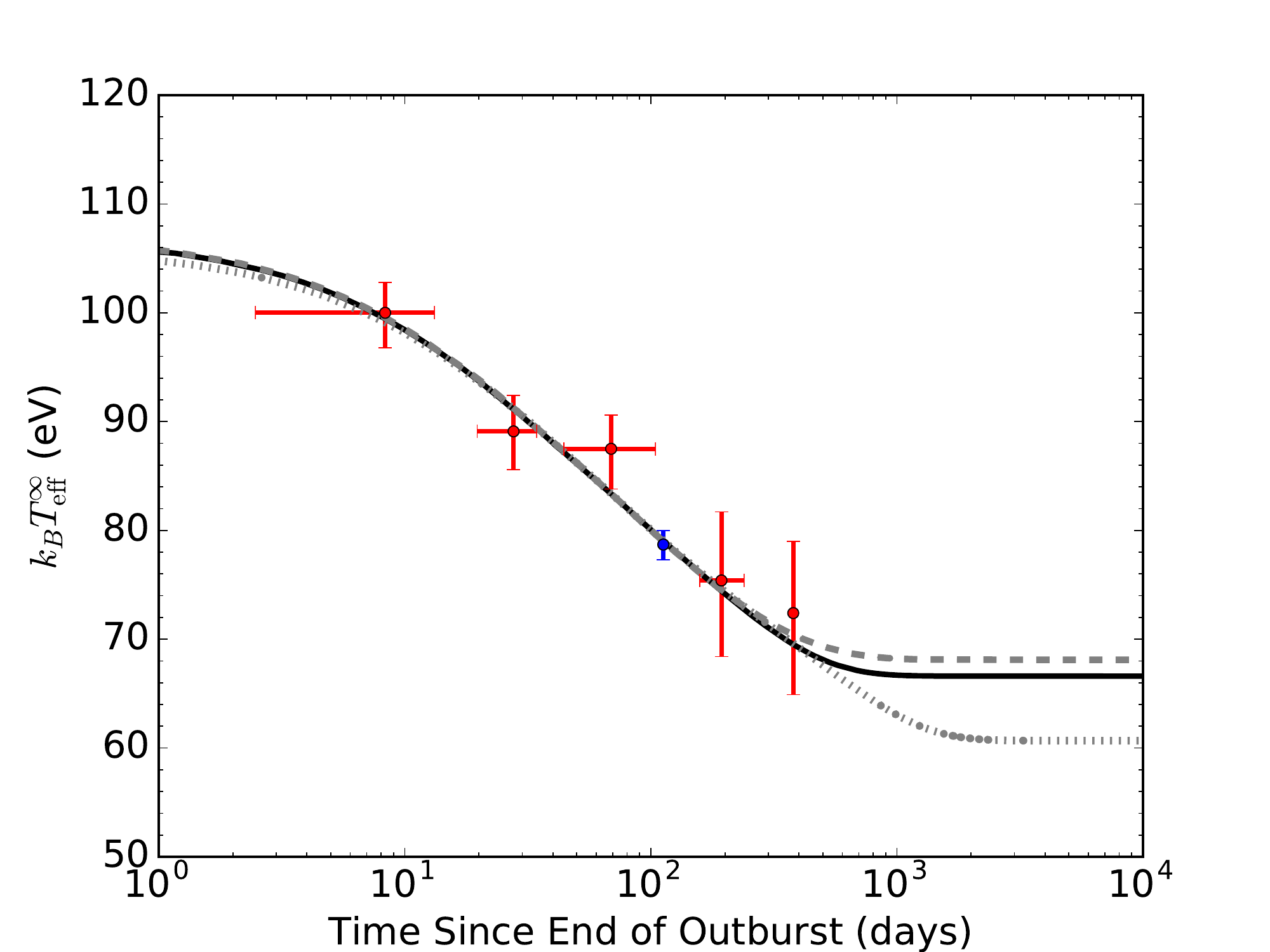}};
\fill[white] (-4.5,1) rectangle (-3.8,-1);
\node at (-4,0) [draw=none, rotate=90]  {$kT^{\infty}_{\mathrm{eff}}$ (eV)};
\end{tikzpicture}
\caption{Quiescent evolution of the effective temperature of 1RXS J1804 after its 2015 outburst. The temperatures determined from {\it Swift}/XRT data are shown in red and the {\it XMM-Newton} data in blue. The solid black line shows the fit using \texttt{NSC\textsc{ool}} model A using a single $Q_\mathrm{imp}$, the dotted grey line shows model B which uses different $Q_\mathrm{imp}$ values in the inner and outer crust, and the dashed grey line indicates model C for which the deep crustal heating has been turned off. The \texttt{NSC\textsc{ool}} fit only accounts for the error on the temperatures and not on time. The errors are in the 90$\%$ confidence range.}
\label{img_cooling_1rxs}
\end{figure}

\subsection{\textit{\textbf{Chandra}}}
The average count rate for the {\it Chandra} observation of the source was (3.6 $\pm$ 0.5)$\times10^{-2}$ counts s$^{-1}$. The spectrum was extracted by the meta-task \texttt{specextract} and grouped to have at least 5 photons per bin using \texttt{grppha}. \texttt{XS\textsc{pec}} was used to fit the spectrum, employing  W-statistics, and the $N_{\mathrm{H}}$ was fixed again to 0.4$\times 10^{22}$ cm$^{-2}$. A simple power-law component fit indicated a soft spectra having a photon index of $\Gamma\sim$2.8. A single thermal component \texttt{nsatmos} model did not fit the data well. An additional power-law component was required for this observation as well but the photon index could not be well constrained. The combined \texttt{nsatmos} and \texttt{powerlaw} fit indicated an unabsorbed 0.5 -- 10 keV flux of $F$ = (5.4 $\pm$ 0.8)$\times$ 10$^{-13}$ erg cm$^{-2}$ s$^{-1}$ ($L_{\mathrm{X}}$ = 21.8 $\pm$ 3.1$\times 10^{32}$ erg s$^{-1}$).


\subsection{Modeling the observed thermal evolution}
\label{sect_crust_cooling_model}
As 1RXS J1804 evolved further into quiescence its spectra continued to soften. When fitting the spectra with an \texttt{nsatmos} model we found a gradual decay in the inferred neutron star surface temperature $kT^{\infty}_{\mathrm{eff}}$, as shown in Table \ref{obsID_grouping} and Figure \ref{img_cooling_1rxs}. The $kT^{\infty}_{\mathrm{eff}}$ decreased from $\sim$100 to $\sim$71 eV between $\sim$8 to $\sim$379 days of quiescence. 

The temperature decrease in Figure \ref{img_cooling_1rxs} might suggest crust cooling of the neutron star in 1RXS J1804 which had been heated out of equilibrium during its 2015 outburst.  Thus, we model the thermal evolution of the neutron star crust during the accretion episode and subsequent cooling period using the crust cooling code \texttt{NSC\textsc{ool}} \citep{page2013forecasting}, taking into account the variability of the mass accretion rate during outburst from the observed light curve \citep{ootes2016}. This accretion rate variability needs to be accounted for as changes in the outburst affects the temperature profile in the crust. The knowledge of the temperature profile in the crust before quiescence is crucial for interpreting the quiescent cooling curve trend \citep{ootes2016}. Comparing the detected surface temperatures with the calculated cooling curves allows us to constrain the parameters of the neutron star crust in 1RXS J1804.

The daily averaged accretion rate during outburst (2015 January 19 to 2015 June 6) was determined from the count rates obtained using {\it Swift}/XRT (0.5 -- 10 keV) and {\it MAXI} (2 -- 10 keV, see \citeauthor{degenaar2016disk} \citeyear{degenaar2016disk} for the {\it MAXI} light curve). We combined data from these two different telescopes to obtain the best constraints on the outburst variability. {\it Swift}/XRT count rates were given preference if data were available from both instruments on the same day. The outburst profile was modelled in terms of bolometric flux. Thus, to transform count rates to bolometric flux a conversion factor was required for {\it Swift} and {\it MAXI}. 1RXS J1804 showed a state transition from hard to soft around 2015 April 3 \citep{degenaar2015neutron,degenaar2016disk} so separate conversion factors were determined for the hard and soft state data. These conversion factors were obtained by dividing the flux as reported by \citet{ludlam2016nustar} in the hard state and \citet{degenaar2016disk} in the soft state by the {\it Swift} or {\it MAXI} count rates that were obtained closest in time to those reported fluxes. We assumed that the fluxes reported by \citeauthor{ludlam2016nustar} \citeyear{ludlam2016nustar} (0.45 -- 50 keV; $F$ = 1.7$\times$10$^{-9}$ erg cm$^{-2}$ s$^{-1}$) and \citeauthor{degenaar2016disk} \citeyear{degenaar2016disk} (0.7 -- 35 keV; $F$ = 4.8$\times$10$^{-9}$ erg cm$^{-2}$ s$^{-1}$) were the bolometric fluxes $F_{\mathrm{bol}}$. The accretion rate was determined using $$\dot{m} = \frac{F_{\mathrm{bol}} 4 \pi D^2}  {\eta_\mathrm{Edd}\, c^2}$$ where $D$ is the distance, $\eta_\mathrm{Edd}$ is the fraction of accreted mass that produces the X-ray luminosity (here we assume $\eta_\mathrm{Edd}$ = 0.2), and $c$ is the speed of light. Using those conversion factors we could also calculate the average accretion rate during outburst (using the profile as shown in Figure \ref{img_1rxs_lightcurve}) and we found that to be $\langle \dot{M} \rangle$ = 0.034 $\dot{M}_\mathrm{Edd}$ (where $\dot{M}_\mathrm{Edd}$ = 1.72$\times 10^{18}$ g s$^{-1}$).

The mass of the neutron star was fixed to $M_{\mathrm{NS}} = 1.6 \,M_\odot$
and the radius to $R_{\mathrm{NS}} = 11$ km in our models, to be consistent with the spectral analysis.
The remaining input parameters were adjusted by hand until a best-fit of the observed surface
temperatures was obtained. These parameters included the core temperature prior to the accretion episode ($T_0$), the amount of shallow heat released per accreted nucleon ($Q_\mathrm{sh}$), the density at which the shallow heating is injected ($\rho_\mathrm{sh}$), 
the depth of the light elements layer in the neutron star envelope ($y_\mathrm{light}$), 
and the impurity parameter of the crust ($Q_\mathrm{imp}$). We tried models with (A) a uniform $Q_\mathrm{imp}$ in the whole crust and with (B) two values, a high $Q_\mathrm{imp}$ in the outer crust and a lower $Q_\mathrm{imp}$ in the inner crust  as proposed by \citeauthor{page2013forecasting} (\citeyear{page2013forecasting}; modelled having a smooth $Q_\mathrm{imp}$ function). We also probed a third family of models with (C) using a uniform $Q_\mathrm{imp}$ but where the deep crustal heating has been turned off and the only heating is provided by the shallow heating. This is to test for the possibility that the deep crustal heating processes are not active or only present at a much reduced level than what is usually assumed (e.g., due to the presence of an hybrid crust as proposed by \citeauthor{wijnands2013testing} \citeyear{wijnands2013testing}). We refer to \citet{page2013forecasting} and \cite{ootes2016} for full details of the crust cooling model. The best fit parameters for these three families of models are presented in Table \ref{tab_fit_NSCool} and Figure \ref{img_cooling_1rxs}.

\begin{table}
\centering
\caption{Fit parameters from \texttt{NSC\textsc{ool}}. The mass of the neutron star was fixed to $M_{\mathrm{NS}} = 1.6\,M_\odot$ and the radius to $R_{\mathrm{NS}} = 11$ km.}
\label{tab_fit_NSCool}
\begin{tabular}{llllll}
\hline
Model & $T_{0}$ &$y_\mathrm{light}$ & $Q_{\mathrm{sh}}$& $\rho_{\mathrm{sh, min}}$& $Q_\mathrm{imp}$\tabularnewline
 & (K) &(g cm$^{-2}$)  & (MeV) & (g cm$^{-3}$)& \tabularnewline
\hline 
A & 4.9$\times 10^{7}$ & 1$\times10^8$& 1.86 & 4$\times 10^{8}$&1 \tabularnewline
B & 4.0$\times 10^{7}$ & 1$\times10^8$& 1.51 & 4$\times 10^{8}$&30 -- 1 \tabularnewline
C & 5.2$\times 10^{7}$ & 1$\times10^8$& 1.90 & 4$\times 10^{8}$&1 \tabularnewline
\hline
\end{tabular}
\end{table}

\FloatBarrier
\section{Discussion}
We monitored the neutron star LMXB 1RXS J1804 after its 2015 outburst with {\it Swift} and {\it XMM-Newton} to search for the cooling of an accretion-heated neutron star crust. The observations analysed covered $\sim$8 to $\sim$379 days after the estimated transition to quiescence. We also analysed the archived 2011 {\it Chandra} observation to check if this observation could be used to obtain an estimate of the quiescent base level of the source. 

\subsection{Crust Cooling}
\label{sect_crust_cooling}
Initial studies of cooling neutron star LMXB crusts focussed only on sources that had outburst times $\gtrsim$ 1 yr \citep[e.g.,][]{wijnands2002xmm,cackett2008cooling}. But more recent work shows that crust cooling can also be observed in relatively bright X-ray transients that have short outburst times of weeks to months (e.g., IGR J17480$-$2446, Swift J174805.3$-$244637, and Aql X$-$1; \citeauthor{degenaar2013continued} \citeyear{degenaar2013continued}, \citeyear{degenaar2015neutron}, \citeauthor{waterhouse2016constraining} \citeyear{waterhouse2016constraining}). Therefore,  one might expect that crust cooling could also potentially be observed for the neutron star in 1RXS J1804 since it had a relatively bright and intermediately long outburst.

The inferred neutron star surface temperature of 1RXS J1804 indeed shows a gradual decay in quiescence from $\sim$100 to $\sim$71 eV over $\sim$370 days when we fitted the spectra with an \texttt{nsatmos} model. This is consistent with a cooling neutron star crust after the cessation of a $\sim$4.5 month accretion-heating episode. The $kT^{\infty}_\mathrm{eff}$ at the beginning of the possible cooling phase is quite similar to what has been observed for other cooling neutron star LMXB sources (see Figure 5 of \citeauthor{homan2014strongly} \citeyear{homan2014strongly} for details). 
The inferred cooling ($e$-folding) timescale for 1RXS J1804, found by fitting the cooling curve with an exponential decay function plus a constant, is $\sim$76$^{+94}_{-47}$ days. Compared to other sources, as summarised by \citet{homan2014strongly}, 1RXS J1804 has a smaller $e$-folding time. However, \citet{homan2014strongly} report the $e$-folding time of quasi-persistent sources (with outburst durations $\gtrsim$1 yr) whereas 1RXS J1804 experienced a relatively shorter $\sim$4.5 month outburst, more akin to IGR J17480$-$2446, Swift J174805.3$-$244637, and Aql X$-$1. The $e$-folding time of IGR J17480$-$2446, of 157$\pm$62 days, is much larger than 1RXS J1804 (i.e., similar to that of some of the quasi-persistent sources) but that for Swift J174805.3$-$244637, having an $e$-folding time of 77.7$\pm$49.1 days, is much closer to 1RXS J1804 \citep{degenaar2013continued,degenaar2015neutron}. Therefore, no firm conclusions can be drawn from these $e$-folding times.


The theoretical model fit using \texttt{NSC\textsc{ool}}, taking into account the variability of the mass accretion rate during outburst, reveals that the source needs $\sim$1.5 to $\sim$1.9 MeV of shallow heating per accreted nucleon to explain the observed cooling curve, the lowest value being allowed when permitting a low conductivity, i.e., a large $Q_\mathrm{imp}$ (a value of 30 was used; Table \ref{tab_fit_NSCool}), in the outer crust.
As in the other sources where the shallow heating is needed, it is required to reproduce the observed high $kT^{\infty}_\mathrm{eff}$ during the early phase of the post-outburst cooling. This amount of shallow heating is consistent with typical values found for most of the other sources that require it to explain their observations (e.g., \citeauthor{brown2009mapping} \citeyear{brown2009mapping}; \citeauthor{degenaar2011evidence} \citeyear{degenaar2011evidence}; \citeauthor{page2013forecasting} \citeyear{page2013forecasting}; \citeauthor{waterhouse2016constraining} \citeyear{waterhouse2016constraining}; we note that for MAXI J0556$-$332 a much larger amount of shallow heating was necessary to model its crust cooling curve; \citeauthor{deibel2015strong} \citeyear{deibel2015strong}). Our fit model A indicates that  the crust of 1RXS J1804 has a high thermal conductivity with a low impurity content ($Q_\mathrm{imp} =1$), which is similar to what has been found for most of the other sources \citep[e.g.,][]{shternin2007neutron,brown2009mapping,page2013forecasting,ootes2016} but a low conductivity in the outer crust is also a possibility as in our model B, in agreement with \citet{page2013forecasting}. Finally, our model C shows that the currently available cooling curve of 1RXS J1804 can be modelled with no, or only little, deep crustal heating occurring during outburst. This might for example be a possibility if the neutron star in this source would have a hybrid crust consisting partly of accreted material and partly original matter \citep[see][]{wijnands2013testing}. Since deep crustal heating releases most of its energy in the inner crust by pycnonuclear reactions its absence/presence is mostly felt in the late time cooling curve: unfortunately our last two {\it Swift} data points have too large errors to allow detection of such a small difference. However, if 1RXS J1804 is observed further in quiescence then the difference between a cold/warm inner crust due to the absence/presence of deep crustal heating may be observable.

As can be seen from Figure \ref{img_cooling_1rxs}, the \texttt{NSC\textsc{ool}} models A and C suggest that the temperature at the end of our cooling curve might have levelled off (i.e., reach a base-level), indicating that the crust and core might soon be in equilibrium again. However, we note that such behaviour is typical for these kind of simulations and it has already been shown several times that sources continued to cool to temperatures below previously determined base-levels \citep[e.g.,][]{shternin2007neutron,brown2009mapping,fridriksson2010rapid,fridriksson2011variable,cackett2013change}. The current simulations were constructed with input parameters based on neutron star properties that have been inferred from other cooling studies and they can accurately represent the observations.  Therefore, despite the above mentioned uncertainty in determining the base-level, it could still indicate that the source is close to having a crust that is in equilibrium with the core again. If so, observations (i.e., performed using {\it Chandra} or {\it XMM-Newton} to reach the required sensitivity) obtained within the next $\sim$1 -- 2 yr should not show much further temperature evolution. 

Unfortunately the 2011 observation using {\it Chandra} did not allow us to obtain an estimate of the the base quiescent level of 1RXS J1804. The flux from the 2011 {\it Chandra} observation is similar to the {\it Swift} observations considered in interval 1 (in Table \ref{obsID_grouping}). Similar to our {\it XMM-Newton} spectrum, the {\it Chandra} spectrum required a power-law component in addition to the thermal component to obtain a good fit indicating possible low-level accretion during the {\it Chandra} observation as well (see Section \ref{sect_low_level_accr} for the possibility of low-level accretion in 1RXS J1804). Therefore, the {\it Chandra} observation cannot be used to determine (even an upper limit on) the core temperature of the neutron star in 1RXS J1804.

More {\it Swift}, {\it XMM-Newton}, and {\it Chandra} observations will be useful to further constrain the thermal evolution and our cooling model. {\it Swift} is useful to monitor the source frequently which would allow us to determine whether the source has levelled-off, if multiple pointings indicate the same flux level. This would indicate that the crust is in equilibrium with the core again. Using {\it Swift}/XRT observations, we would also get a broad indication of the temperature evolution although the errors on the temperature will be large. {\it XMM-Newton} and {\it Chandra} give better constraints on the inferred neutron star crust temperature. However, they can observe the source much more infrequently than {\it Swift}, thereby missing the long term source evolution.

\subsection{Low-level accretion?}
\label{sect_low_level_accr}
The thermal evolution of 1RXS J1804 can be interpreted as cooling (see Section \ref{sect_crust_cooling}), and the fact that the {\it Swift} data could be well fitted with a single \texttt{nsatmos} model is consistent with that.  However, the {\it XMM-Newton} data statistically required an additional power-law component to obtain a good model fit. This power-law component was statistically not required by the {\it Swift} data, however, its presence could not be ruled out either (see Section \ref{sect_res_swift}). If a similar power-law component was also present during the {\it Swift} observations it would have a systematic effect on the inferred neutron star surface temperature $kT^{\infty}_{\mathrm{eff}}$. It is difficult to determine the outcome of these systematic effects without extensive assumptions (e.g., about the photon indices as well as the actual strength of the possible power-law components), and a full study of this is beyond this scope of the paper. However, at the end of Section \ref{sect_analysis_powerlaw_res} we have shown that the outcome of the single component fit to the {\it Swift} data and multiple component fit to the {\it XMM-Newton} data can be compared meaningfully.

It is possible that the physical process producing the power-law component was only active at the time of the {\it XMM-Newton} observation. However, we consider this quite unlikely. It is more probable that the power-law component is also present during the {\it Swift} observations but could not be detected. We note that a similar power-law component has been detected during some of the observations of other cooling studies \citep[e.g.,][]{fridriksson2010rapid,fridriksson2011variable,degenaar2011further,degenaar2015neutron,waterhouse2016constraining} so also those studies might be affected by the uncertainties in the nature of the power-law component.

It is unclear what the origin of the power-law component is, but it has been suggested that the power-law component in some quiescent neutron star LMXBs is related to low-level accretion on to the neutron star surface \citep[e.g.,][]{campana1998neutron,rutledge2002variable,cackett2010quiescent,chakrabarty2014hard,d2015radiative,wijnands2015low}. Low-level accretion also contributes a soft component to the spectra \citep{zampieri1995x,campana1998neutron}.

The archival 2011 {\it Chandra} observation indicated a higher luminosity than some of the quiescent {\it Swift} data. It showed a similar type of spectra as that from the {\it XMM-Newton} data, requiring a power-law in addition to the thermal component. However, due to the limiting quality of the data the contribution from the power-law component was unconstraining. The source also showed a sub-luminous outburst in 2012 during which its peak luminosity likely did not exceed a few times 10$^{34}$ erg s$^{-1}$ \citep{chenevez2012integral,kaur2012swift}. This low peak luminosity makes it unclear if this was a real outburst or that it was due to a period of an increased low-level accretion rate. This very-faint outburst combined with the higher flux detected during the {\it Chandra} observation compared to our lowest observed {\it Swift} fluxes, indicates that before the 2015 outburst the source exhibited significant variability in quiescence, possible due to variable low-level accretion.

If the power-law component we see in 1RXS J1804 is due to low-level accretion then this would indicate that at least during the {\it XMM-Newton} observation low-level accretion might have been present. The presence of low-level accretion could significantly alter our interpretation of a cooling crust. In the following we explore different scenarios for the observed slow decay in quiescence assuming low-level accretion occurs.


 

{\it Dominant Low-level Accretion} :
Assuming what we observe is low-level accretion and if low-level accretion dominates in quiescence then it is possible that this accretion is strong enough to significantly heat up the neutron star surface and drive the thermal evolution. However, low-level accretion is expected to be variable in a more stochastic way compared to the smooth decay curve we observe. Such (sometimes strong) random fluctuations have been observed for example for the neutron star LMXB Cen X-4 \citep[e.g.,][]{campana2004variable,cackett2010quiescent} and the black hole LMXB V404 Cyg \citep[e.g.,][]{hynes2004correlated,bradley2007spectrum,hynes2009quiescent} in their quiescent state \citep[although sometimes more structured variability, i.e., in the form of brief faint flares, have also been observed in some systems; e.g.,][]{cackett2010quiescent,cackett2011quiescent,fridriksson2011variable,wijnands2013low,bernardini2013daily,zelati2014year}. But this does not entirely rule out the presence of low-level accretion, as an explanation for what we observe for 1RXS J1804, since its physics is not very well understood and could possibly lead to a smooth decay as well (as indicated by the smooth decay that has been observed for the black hole system SWIFT J1756.9$-$2508 by \citeauthor{padilla2013multiwavelength} \citeyear{padilla2013multiwavelength}). If low-level accretion is dominant then our interpretation of the neutron star cooling in quiescence after an accretion outburst is not valid.


{\it Dominant Thermal Component} : 
It may be that the smooth decay observed from the neutron star is dominated by the cooling and variable low-level accretion occurs but its contribution is always overshadowed by the neutron star cooling and therefore does not contribute significantly to the decay trend. Thus, the cooling neutron star drives the thermal evolution and the results presented in Section \ref{sect_crust_cooling} apply directly.

{\it Both the thermal and power-law components contribute significantly} :
There is also an intermediate scenario in which both the thermal and non-thermal components contribute significantly to the X-ray flux and spectrum (e.g., as has been observed in some systems; \citeauthor{cackett2010quiescent} \citeyear{cackett2010quiescent}; \citeauthor{bahramian2013discovery} \citeyear{bahramian2013discovery};  \citeauthor{campana2014return} \citeyear{campana2014return}; see \citeauthor{wijnands2015low} \citeyear{wijnands2015low} for an in-depth discussion). For our {\it XMM-Newton} quiescent data we observe a power-law contribution of $\sim$30 per cent (for a total luminosity of $L_X$ $\sim$0.9$\times 10^{33}$ ($D$/5.8 kpc)$^{2}$ erg s$^{-1}$) which is compatible with a dominant thermal component in quiescence with continued low-level accretion that also contributes to the thermal evolution \citep{wijnands2015low}. If true, it will remain difficult to disentangle both effects and to determine the exact heating and cooling of the crust in this system due to the accretion of matter during its 2015 outburst.

The above discussion, assuming low-level accretion is present, has so far assumed that the physical process producing the power-law component takes place very close to the neutron star and interacts with its surface \citep[e.g., see the discussions in][]{chakrabarty2014hard,d2015radiative,wijnands2015low}. Alternatively, it is also possible that the power-law component is produced far away from the neutron star and so contributes to the total observed luminosity but does not influence the temperature evolution of the neutron star as it does not interact with the neutron star surface. Processes such as shock interactions in a propeller outflow or pulsar wind happen far away from the neutron star and could possibly produce a power-law component in the observed spectra \citep[e.g.,][]{campana1998neutron,zhang1998spectral}. In this case the gradual thermal decay can be correctly interpreted as cooling in spite of the presence of the power-law component. However the observed luminosity is not only that of the cooling, the power-law component also contributes to the observed luminosity and therefore the exact neutron star surface temperature (and with that the exact heating and cooling of the crust) remains difficult to correctly extract from the data.


\section*{Acknowledgements}

AP, RW, and LO are supported by a NWO Top Grant, Module 1, awarded to RW. ND is supported by an EU Marie Curie Intra-European fellowship under contract no. FP-PEOPLE-2013-IEF-627148 and an NWO Vidi grant. DP is partially supported by the Consejo Nacional de Ciencia y Tecnolog{\'\i}a with a CB-2014-1 grant $\#$240512. DA acknowledges support from the Royal Society. JWTH acknowledges funding from an NWO Vidi fellowship and from the European Research Council under the European Union's Seventh Framework Programme (FP/2007-2013) / ERC Starting Grant agreement nr. 337062 (``DRAGNET"). ML was supported by the Spanish Ministry of Economy and Competitiveness under the grant AYA2013-42627. JCAMJ is supported by an Australian Research Council Future Fellowship (FT140101082). The authors are grateful to Norbert Schartel and the {\it XMM-Newton} scheduling team for performing the DDT observation used in this work. We acknowledge the use of public data from the {\it Swift} data archive.




\bibliographystyle{mnras}

 \newcommand{\noop}[1]{}



%
%


\bsp	
\label{lastpage}
\end{document}